\documentclass[twocolumn]{aastex631} %,linenumbers
\usepackage{multirow}
\usepackage{subfigure}
\usepackage{amsmath}

\begin{document}
\title{Testing the Cosmic Distance Duality Relation Using Strong Gravitational Lensing Time Delays and Type Ia Supernovae}

%\maketitle

\author{Jing-Zhao Qi}
\affiliation{Department of Physics, College of Sciences, Northeastern University, Shenyang 110819, China}

\author{Yi-Fan Jiang}
\affiliation{Department of Physics, College of Sciences, Northeastern University, Shenyang 110819, China}

\author{Wan-Ting Hou}
\affiliation{School of Mathematics and Statistics, Liaoning University, Shenyang 110819, China}

\author{Xin Zhang}
\affiliation{Department of Physics, College of Sciences, Northeastern University, Shenyang 110819, China}
\affiliation{Key Laboratory of Cosmology and Astrophysics (Liaoning Province), Northeastern University, Shenyang 110819, China}
\affiliation{Key Laboratory of Data Analytics and Optimization for Smart Industry (Ministry of Education), Northeastern University, Shenyang 110819, China}
\affiliation{National Frontiers Science Center for Industrial Intelligence and Systems Optimization, Northeastern University, Shenyang 110819, China}

\correspondingauthor{Xin Zhang}
\email{zhangxin@mail.neu.edu.cn}

\begin{abstract}
We present a comprehensive test of the cosmic distance duality relation (DDR) using a combination of strong gravitational lensing (SGL) time delay measurements and Type Ia supernovae (SNe Ia) data. We investigate three different parameterizations of potential DDR violations. To bridge the gap between SGL and SNe Ia datasets, we implement an artificial neural network approach to reconstruct the distance modulus of SNe Ia. Our analysis uniquely considers both scenarios where the absolute magnitude of SNe Ia ($M_B$) is treated as a free parameter and where it is fixed to a Cepheid-calibrated value. Using a sample of six SGL systems and the Pantheon+ SNe Ia dataset, we find no statistically significant evidence for DDR violations across all parameterizations. The consistency of our findings across different parameterizations not only reinforces confidence in the standard DDR but also demonstrates the robustness of our analytical approach.
\end{abstract}

\section{Introduction}
The cosmic distance duality relation (DDR), also known as the Etherington relation \citep{Etherington1933}, is a fundamental concept of observational cosmology. It establishes a connection between the luminosity distance ($D_L$) and the angular diameter distance ($D_A$) in the form of $D_L = (1+z)^2 D_A$, where $z$ is the redshift. The validity of the DDR is a direct consequence of the metric theory of gravity, requiring that the number of photons is conserved along the light path and that photons travel along null geodesics \citep{bassett2004cosmic}. Therefore, testing the validity of the DDR is crucial for verifying the foundations of modern cosmology and the theories of gravity \citep{Avgoustidis2010, Holanda2011}.

On the other hand, the current state of cosmology is facing a severe crisis due to the increasing tension between the measurements of cosmological parameters from the early and late Universe \citep{Planck:2018vyg,Riess:2021jrx,Qi:2020rmm,DiValentino:2019qzk,Handley:2019tkm,Heymans:2020gsg,Zhao:2017urm,Feng:2017mfs,Feng:2019jqa,Gao:2021xnk,Gao:2022ahg,Zhang:2014dxk}. Among these discrepancies, the most significant one is known as the ``Hubble tension", which has been consistently observed in various cosmological probes \citep{di2021realm,Vagnozzi:2019ezj,Zhang:2019ylr,Guo:2018ans,di2021snowmass2021}. The Hubble constant ($H_0$), representing the current expansion rate of the Universe, shows a significant difference between values obtained from early Universe observations, particularly the cosmic microwave background \citep{Planck:2018vyg}, and those from late Universe observations, such as the local distance ladder using Cepheid variables and Type Ia supernovae (SNe Ia; \citealt{Riess:2021jrx,bhardwaj2023high}). This tension has reached a statistical significance of more than 5$\sigma$, indicating a serious discrepancy between the standard cosmological model and observations \citep{di2021realm}.

The persistence of the Hubble tension, despite extensive efforts to resolve it, suggests that there might be new physics beyond the standard cosmological model, or that some of the fundamental assumptions in cosmology may need to be revised \citep{hu2023hubble}. One of these assumptions is the validity of the DDR. If the DDR is violated, it could lead to a miscalibration of the distances measured by different cosmological probes, potentially contributing to the observed tension \citep{Renzi:2020fnx,Renzi:2021xii}. Therefore, testing the validity of the DDR is of utmost importance in the context of the current cosmological crisis.

In recent years, various combinations of cosmological probes have been employed to test the DDR, which requires simultaneous measurements of luminosity distances and angular diameter distances. SNe Ia have been the primary source for luminosity distance measurements, while angular diameter distances have been provided by different cosmological probes. These include baryon acoustic oscillations \citep{Wu:2015ixa,Xu:2020fxj,Favale:2024sdq}, the Sunyaev--Zeldovich effect together with X-ray emission of galaxy clusters \citep{Holanda:2019vmh}, radio quasars \citep{Qi:2019spg,Tonghua:2023hdz,Liu:2021fka} and strong gravitational lensing (SGL) systems \citep{lyu2020testing,Lima:2021slf,Holanda:2021vvh,Liao:2019xug,Rana:2017sfr,Liao:2015uzb}. More recently, gravitational waves have emerged as a promising new standard siren for luminosity distance measurements \citep{Zhang:2019loq,Qi:2021iic,Wang:2021srv,Jin:2023sfc,Hou:2022rvk,Jin:2021pcv,Zhao:2019gyk,Li:2019ajo}, opening up new avenues for DDR testing \citep{Qi:2019spg,Fu:2019oll,Yang:2017bkv}. The combination of these diverse probes allows for robust and multifaceted examinations of the DDR across different cosmic epochs and distance scales. Among these probes, SGL has emerged as a powerful tool for cosmological studies \citep{Wong:2019kwg,Wang:2021kxc}. It can provide independent measurements of both the angular diameter distance and the time-delay distance, offering an absolute distance measurement \citep{Qi:2022kfg}.

SGL occurs when the light from a distant source is deflected by a massive foreground galaxy, resulting in multiple images of the source. The positions and fluxes of these images depend on the mass distribution of the lens galaxy and the distances between the observer, the lens, and the source. In particular, the maximum angular separation between the lensed images has been widely used to test the DDR \citep{Lima:2021slf,Holanda:2021vvh,Rana:2017sfr}. However, this approach assumes a homogeneous mass distribution for all lens galaxies, typically described by a singular isothermal sphere or a singular isothermal ellipsoid model. Such an assumption can potentially introduce biases in the results, as the mass distribution of lens galaxies may not be identical \citep{Qi:2018aio,Wang:2022rvf}.

An alternative observable in SGL systems is the time delay between the lensed images. The time delay arises due to the difference in the path lengths and the gravitational potential traversed by the light rays from the source to the observer \citep{Refsdal:1964nw,Suyu:2009by}. Measuring the time delays requires long-term monitoring of the lensed system, spanning several years or even decades \citep{Wong:2019kwg}. Consequently, the sample size of SGL time delay measurements is currently limited, with only seven well-studied systems available \citep{Wong:2019kwg}. However, each of these systems has been modeled individually, accounting for the specific mass distribution of the lens galaxy \citep{Wong:2019kwg}. This approach makes the time delay measurements more accurate and less prone to biases compared to the angular separation method.

Despite the advantages of using SGL time delays for cosmological tests, the independent modeling of each system introduces additional complexities in the analysis. When combining SGL time delay data with other cosmological probes, such as SNe Ia, to test the DDR, the uncertainties from both datasets need to be properly accounted for, which can be a challenging task. Nevertheless, the precision offered by SGL time delay measurements makes them a valuable tool for testing the DDR and exploring potential deviations from the standard cosmological model.

In this paper, we present a comprehensive analysis of the DDR using a combination of SGL time delay measurements and SNe Ia data. We aim to employ the strengths of both datasets to provide a robust test of the DDR and to investigate any potential violations of this fundamental relation. 

The paper is organized as follows: Section \ref{sec:data} describes the SGL time delay and SNe Ia datasets used in our analysis. We also outline the methodology employed to test the DDR, including the distance reconstruction method and the parameterization of potential DDR violations. In Section \ref{sec:results}, we present the results and discuss their implications for the validity of the DDR and the underlying theories of gravity. Finally, we summarize our findings and provide an outlook for future work in Section \ref{sec:conclusion}.

\section{Data and Methodology}\label{sec:data}

\subsection{Strong Gravitational Lensing Time Delays}

SGL occurs when light rays from a distant source are deflected by an intervening massive object, creating multiple images of the source. The travel time of light from the source to the observer depends on both the path length and the gravitational potential it traverses. For a single lens plane, we can quantify the extra time a light ray takes compared to an unlensed path using the excess time delay function \citep{Wong:2019kwg}:
\begin{equation}
t(\boldsymbol{\theta}, \boldsymbol{\beta}) = \frac{D_{\Delta t}}{c} \left[\frac{(\boldsymbol{\theta} - \boldsymbol{\beta})^2}{2} - \psi(\boldsymbol{\theta})\right],
\end{equation}
where $\boldsymbol{\theta}$ represents the image position in the lens plane, $\boldsymbol{\beta}$ is the source position, $D_{\Delta t}$ is the time-delay distance, and $\psi(\boldsymbol{\theta})$ denotes the lens potential. The time-delay distance is a composite distance measure defined as \citep{Refsdal:1964nw,Schneider:1992bmb,Suyu:2009by}
\begin{equation}
D_{\Delta t} \equiv (1 + z_d) \frac{D_d D_s}{D_{ds}},
\end{equation}
where $z_d$ is the lens redshift, and $D_d$, $D_s$, and $D_{ds}$ are the angular diameter distances to the lens, to the source, and between the lens and source, respectively.

In systems with multiple images, we can measure the time delay $\Delta t_{ij}$ between two images $i$ and $j$ \citep{Wong:2019kwg}:
\begin{equation}
\Delta t_{ij} = \frac{D_{\Delta t}}{c} \left[\frac{(\boldsymbol{\theta}_i - \boldsymbol{\beta})^2}{2} - \psi(\boldsymbol{\theta}_i) - \frac{(\boldsymbol{\theta}_j - \boldsymbol{\beta})^2}{2} + \psi(\boldsymbol{\theta}_j)\right].
\end{equation}
This equation encapsulates the difference in excess time delays between the two images.

By carefully monitoring the flux variations of multiple images and constructing accurate lens models, the time delays and $D_{\Delta t}$ can be measured. This approach provides a unique method for probing cosmological parameters, especially $H_0$, independent of other techniques like the cosmic distance ladder.

In our work, seven systems were considered initially. However, one system with a redshift of 2.375 was excluded as it exceeded the maximum redshift (\textit{z} = 2.26) of the SNe Ia sample, making direct calibration impossible. Our analysis utilizes a carefully selected sample of six well-studied SGL systems with precisely measured time delays between the lensed images released by the H0LiCOW collaboration: B1608+656 \citep{Suyu:2009by, Jee:2019hah}, RXJ1131-1231 \citep{Chen:2019ejq, Suyu:2012aa}, HE 0435-1223 \citep{Wong:2016dpo, Chen:2019ejq}, SDSS 1206+4332 \citep{Birrer:2018vtm}, WFI2033-4723 \citep{Rusu:2019xrq}, and PG 1115+080 \citep{Chen:2019ejq}. 

These SGL systems have been meticulously monitored over several years, resulting in high-precision time delay measurements \citep{Wong:2019kwg}. A key strength of this dataset is that each system has been individually modeled, taking into account the specific mass distribution of the lens galaxy \citep{Wong:2019kwg}. This tailored approach significantly reduces potential biases in the measurements and provides robust estimates of both the time delays and their associated uncertainties.

Integrating the SGL data with the SNe Ia dataset for the DDR test presents a challenge due to the complex nature of the SGL time delay likelihood functions. Each SGL system has its own distinct likelihood, making the consideration of SNe Ia uncertainties difficult. To address this, we developed a modified version of the likelihood analysis based on the publicly available H0LiCOW code\footnote{\url{https://github.com/shsuyu/H0LiCOW-public}}. Our adaptations to this code allowed for the appropriate integration of SNe Ia error contributions into the SGL time delay likelihoods. The theoretical formulation of these modified likelihood functions is detailed in Appendix \ref{app} of this paper, providing a transparent and reproducible methodology for our analysis.

\subsection{Type Ia Supernovae}
In this paper, we employ the SNe Ia Pantheon+ compilation, which contains 1701 light curves of 1550 unique SNe Ia in the redshift range $0.001 < z < 2.26$ \citep{Brout:2022vxf}, to test the DDR. Compared to the original Pantheon compilation \citep{Pan-STARRS1:2017jku}, the sample size of the Pantheon+ compilation has greatly increased, and the treatments of systematic uncertainties in redshifts, peculiar velocities, photometric calibration, and intrinsic scatter model of SNe Ia have been improved.

It should be noted that not all the SNe Ia in the Pantheon+ compilation are included in the Pantheon+ compilation. Since the sensitivity of peculiar velocities is very large at low redshift ($z < 0.008$), which may lead to biased results, we adopt the processing treatments used by \citet{Brout:2022vxf}, namely, removing the data points in the redshift range $z < 0.01$.

For each SN Ia, the observed distance modulus is given by
\begin{equation}
\mu_{\mathrm{SN}}=m_B-M_B, \label{con:E2.1}
\end{equation}
where $m_B$ is the observed magnitude in the rest-frame \textit{B}-band. The theoretical distance modulus $\mu_{\mathrm{th}}$ is defined as
\begin{equation}
\mu_{\mathrm{th}}=5\log_{10}\left[\frac{D_L(z)}{\mathrm{Mpc}}\right]+25, \label{con:E2.2}
\end{equation}
where ${D_L(z)}$ is the luminosity distance.

\subsection{Reconstruction Method: Artificial Neural Network}
To reconstruct the distance modulus $m_B$ of SNe Ia, we employ a nonparametric reconstruction technique based on artificial neural networks (ANN). The ANN method, implemented using the \texttt{REFANN} \citep{Wang:2019vxv} Python code, allows us to reconstruct a function from data without assuming a specific parameterization. This approach has been widely applied in various cosmological studies \citep{Qi:2023oxv,Dialektopoulos:2021wde,Benisty:2022psx}.

\texttt{REFANN} offers several key advantages that are crucial for our analysis of the cosmic distance duality relation. As a data-driven approach, \texttt{REFANN} can reconstruct the distance--redshift relation without assuming any specific functional form, allowing us to capture potential anomalies or subtle features that might indicate violations of the distance duality relation. It provides robust error estimation at interpolated points, which is essential for our analysis of potential DDR violations.

While other nonparametric reconstruction methods, such as Gaussian process (GP) regression, have been widely used in cosmological studies \citep{Seikel:2012uu, Seikel:2012cs,Benisty:2022psx,Qi:2018aio,Qi:2023oxv}, \texttt{REFANN} offers unique benefits for our specific research goals. Unlike GP, which requires the specification of a kernel function and can be computationally intensive for large datasets, \texttt{REFANN} adapts more flexibly to the underlying data structure without such constraints. Furthermore, \texttt{REFANN} has demonstrated superior performance in handling non-Gaussian errors and capturing complex, nonlinear relationships in cosmological data, which are crucial aspects of our DDR analysis.

\texttt{REFANN}'s flexibility in handling complex data relationships and its proven effectiveness in cosmological studies make it particularly well-suited for our research goals. These characteristics enable us to perform a more comprehensive and reliable test of the cosmic distance duality relation compared to traditional smoothing methods and other nonparametric approaches.

The optimal ANN model used for reconstructing $m_B$ consists of one hidden layer with 4096 neurons in total. By training the ANN on the Pantheon+ dataset, we obtain a data-driven reconstruction of the distance modulus as a function of redshift. This reconstruction enables us to directly extract the luminosity distance $D_L(z)$ at the redshift of each strong gravitational lensing sample, facilitating a direct comparison with the angular diameter distances $D_A(z)$ derived from the lensing observations.

The ANN reconstruction of $m_B$ provides a flexible and model-independent approach to estimate the luminosity distances of SNe Ia. By using the expressive power of neural networks, the ANN method can capture complex nonlinear relationships between the distance modulus and redshift, allowing for a more accurate representation of the underlying cosmological distance--redshift relation.

Furthermore, the training and optimization processes of ANN can introduce additional uncertainties and sensitivities. The choice of hyperparameters, such as the network architecture and training strategy, can impact the flexibility and generalization capabilities of the ANN model. These factors may contribute to the observed variations in the reconstructed distance moduli and the associated confidence regions.

\begin{figure}
\centering
\includegraphics[scale=0.45]{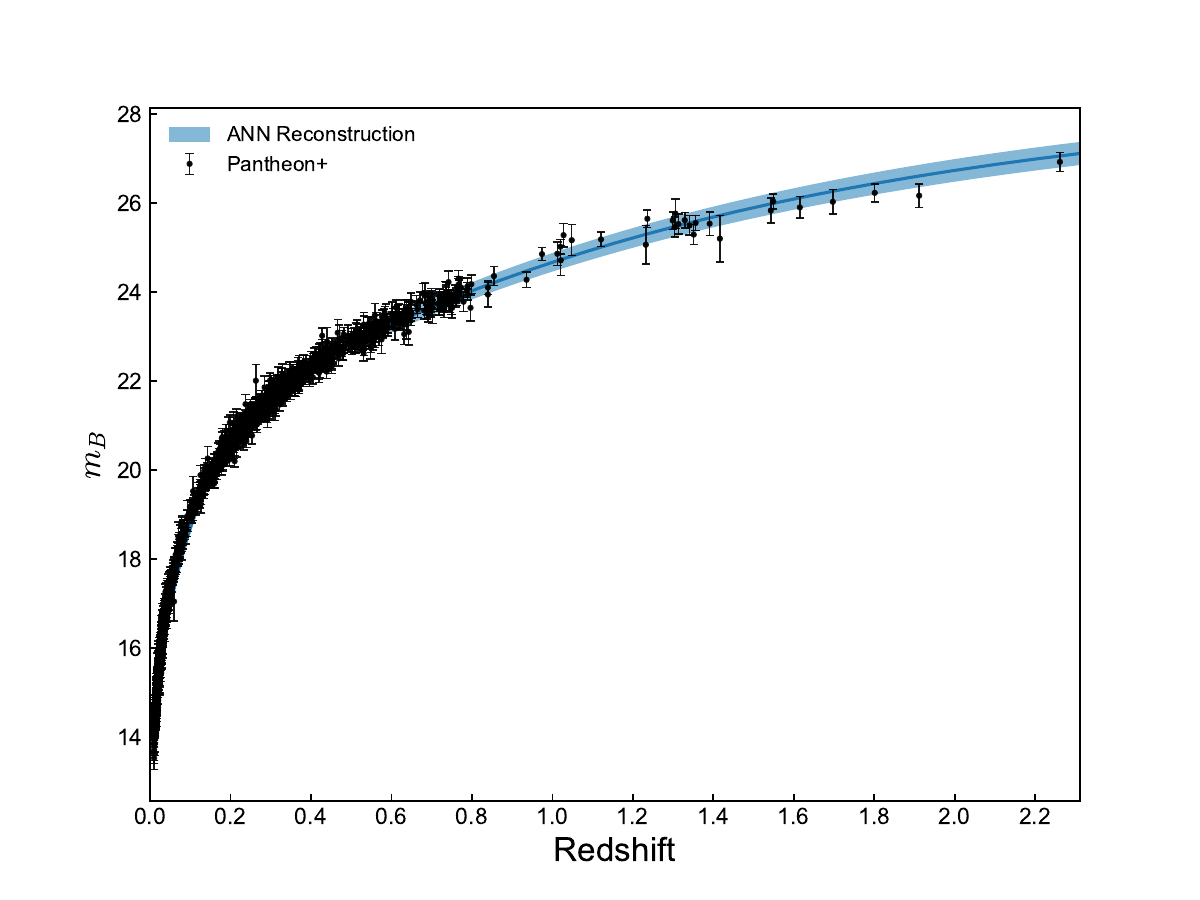}
\caption{Comparison of the distance modulus ($m_B$) reconstruction using ANN with the original Pantheon+ dataset across redshift. The blue shaded area represents the 1$\sigma$ confidence region of the ANN reconstruction, while the black points with error bars show the individual Pantheon+ data points. The ANN reconstruction closely follows the trend of the observational data, providing a smooth and continuous representation of the distance modulus--redshift relation up to $z \approx 2.3$.}
\label{fig:fig1}
\end{figure}

Despite these challenges, the ANN reconstruction method provides a valuable tool for testing the DDR using SNe Ia and strong gravitational lensing data. By reconstructing the distance modulus $m_B$ in a nonparametric method, we can obtain a model-independent estimate of the luminosity distances, which can be directly compared with the angular diameter distances derived from lensing observations. This approach allows us to probe the validity of the DDR and explore potential deviations from the standard cosmological model without relying on specific parametrizations or assumptions about the underlying cosmology. The reconstruction of the distance modulus $m_B$ is shown in Figure \ref{fig:fig1}.

\subsection{Parameterization of DDR Violations}

To investigate potential violations of the DDR, we introduce a parameterization that allows for deviations from the standard relation,
\begin{equation}
    D_L D_A^{-1} (1+z)^{-2} = \eta(z),
\end{equation}
where $\eta(z)$ quantifies the deviation from the standard DDR. We consider three different parameterizations for $\eta(z)$ \citep{Yang:2017bkv}:
\begin{eqnarray}
    P_1: \eta(z) &=& 1 + \eta_0 z,\\
    P_2: \eta(z) &=& 1 + \eta_0 \frac{z}{1+z},\\
    P_3: \eta(z) &=& 1 + \eta_0 \log(1+z),
\end{eqnarray}
where $\eta_0$ is a constant that quantifies the magnitude of the potential DDR violation. In the standard cosmological model, $\eta_0 = 0$, and any statistically significant deviation from zero would indicate a violation of the DDR.

With the reconstructed $m_B$ by ANN method, we extract the $m_B$ values at the redshifts corresponding to the SGL systems and convert them to luminosity distances ($D_L$), associating the $\eta(z)$ parameter and the absolute magnitude of SNe Ia ($M_B$). This allows us to directly compare the $D_L$ values derived from SNe Ia with the angular diameter distances ($D_A$) obtained from the SGL measurements.

Using the $D_L$ obtained from supernova observations, we derive the $D_A$ by applying DDR with a potential violation parameter $\eta(z)$:
\begin{equation}
D_A(z) = D_L(z) / [(1+z)^2 \eta(z)]
\end{equation}

For each strong lensing system, we use these converted $D_A$ values to calculate the angular diameter distances to the lens ($D_d = D_A(z_d)$) and to the source ($D_s = D_A(z_s)$). We then compute the ratio of the lens-source distance to the source distance ($D_{ds} / D_s$) using the equation
\begin{equation}
D_{ds}/D_s = 1 - [(1+z_d)D_d] / [(1+z_s)D_s]
\end{equation}
Finally, we calculate a theoretical time-delay distance ($D_{\Delta t}$) using
\begin{equation}
D_{\Delta t} = (1+z_d) (D_d D_s) / D_{ds}
\end{equation}
This derived $D_{\Delta t}$ is then compared with the observed time-delay distance from strong lensing measurements. The detailed mathematical formulations for converting luminosity distances to angular diameter distances and time-delay distances, as well as the associated error propagation calculations, are provided in Appendix \ref{app:distance_conversion}. These equations form the foundation of our analysis, allowing us to test the cosmic distance duality relation using the combination of strong gravitational lensing time delays and SNe Ia data.

\section{Results and Discussion}\label{sec:results}
To constrain $\eta_0$, we employed the Python package \texttt{emcee} \citep{Foreman-Mackey:2012any} to implement a Markov Chain Monte Carlo (MCMC) analysis. Our approach maximized the modified likelihood function detailed in Appendix \ref{app}, which integrates both the SNe Ia and SGL time delay observational data, yielding robust constraints on the cosmic distance duality relation. Our findings are presented in two parts: first, we discuss the results when the absolute magnitude of SNe Ia ($M_B$) is allowed to vary, and then we examine the case where $M_B$ is fixed to a specific value. All of the constraint results for $\eta_0$ and $M_B$ are listed in Table \ref{tab:results}.

\begin{table}[htbp]
\centering
\caption{Constraints on $\eta_0$ and $M_B$ for different parameterizations}
\label{tab:results}
\begin{tabular}{lccc}
\hline
Model & $\eta_0$ & $M_B$ & Scenario \\
      &          & (mag) &           \\
\hline
P1  & $0.20^{+0.11}_{-0.21}$ & $-18.86^{+0.21}_{-0.56}$ & ... \\
P2  & $0.19^{+0.17}_{-0.23}$ & $-19.00^{+0.22}_{-0.56}$ & $M_B$ free \\
P3  & $0.17^{+0.18}_{-0.23}$ & $-19.08^{+0.31}_{-0.49}$ & ... \\
\hline
P1  & $-0.02 \pm 0.13$  & $-19.253 \pm 0.027$ & ... \\
P2  & $-0.03^{+0.23}_{-0.54}$ & $-19.253 \pm 0.027$ & $M_B$ fixed \\
P3  & $-0.19^{+0.24}_{-0.37}$ & $-19.253 \pm 0.027$ & ... \\
\hline
\end{tabular}
\end{table}

\subsection{Analysis with Varying Absolute Magnitude of SNe Ia}

\begin{figure*}
\centering
\includegraphics[scale=0.35]{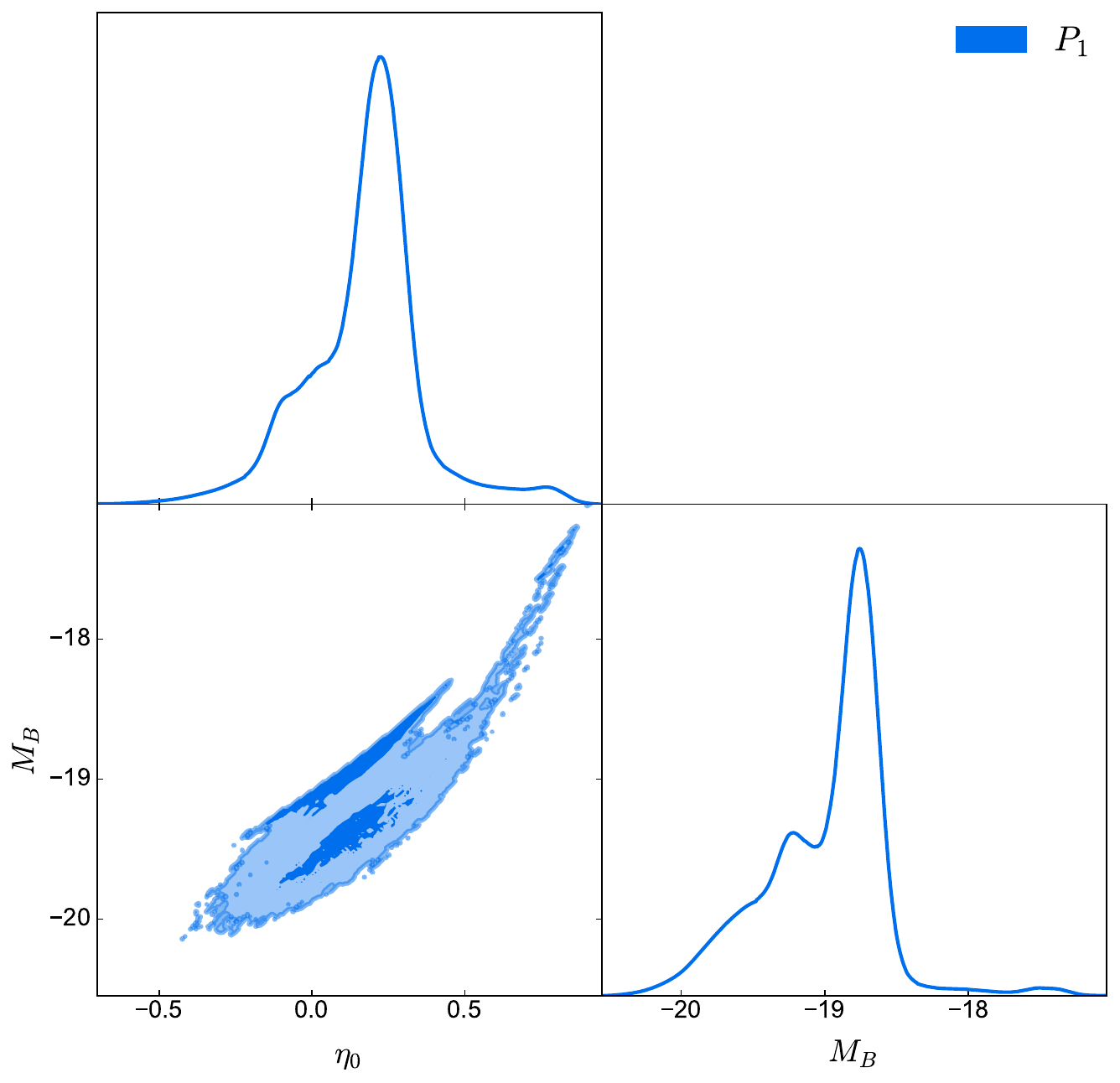}
\includegraphics[scale=0.35]{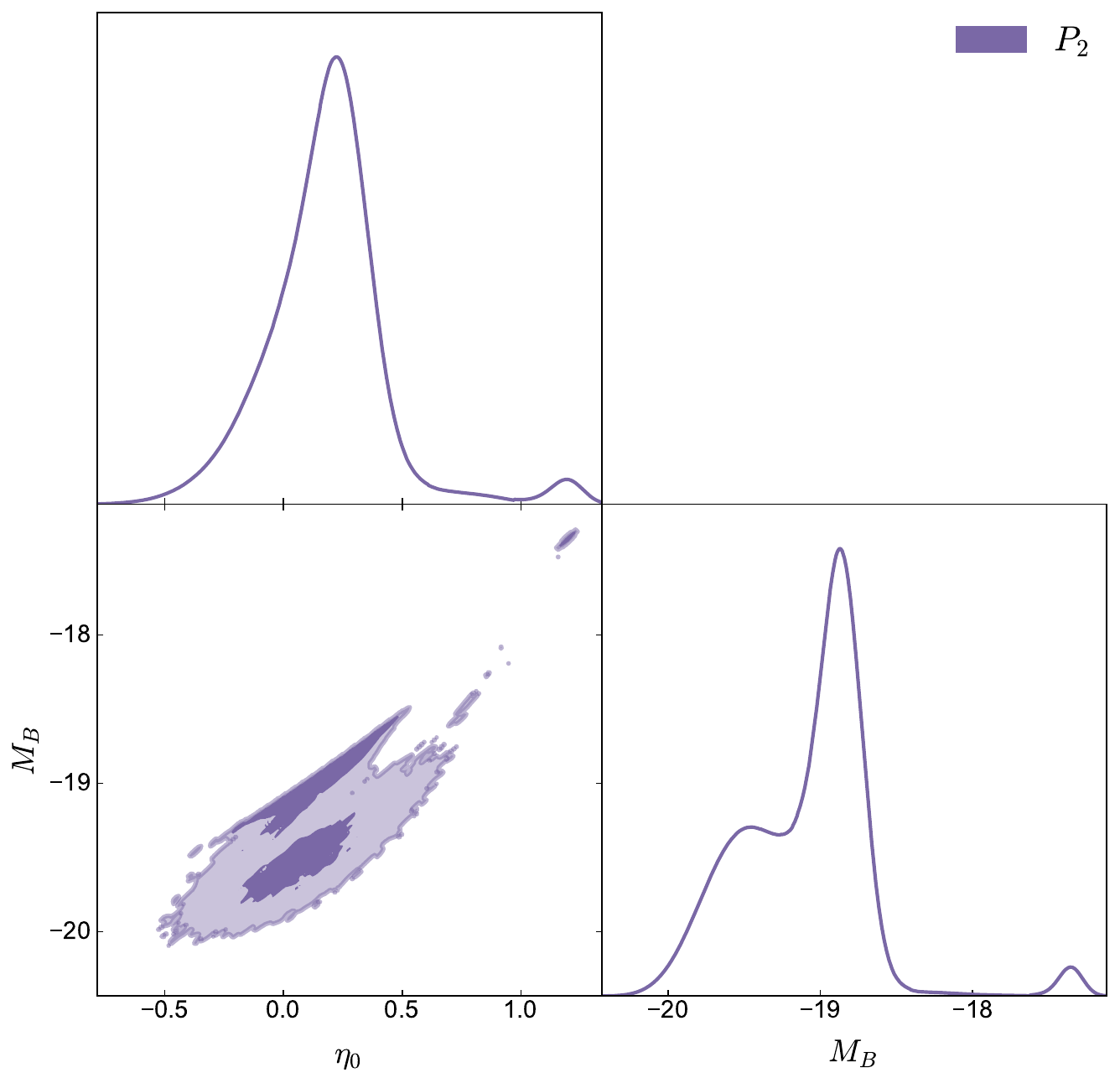}
\includegraphics[scale=0.35]{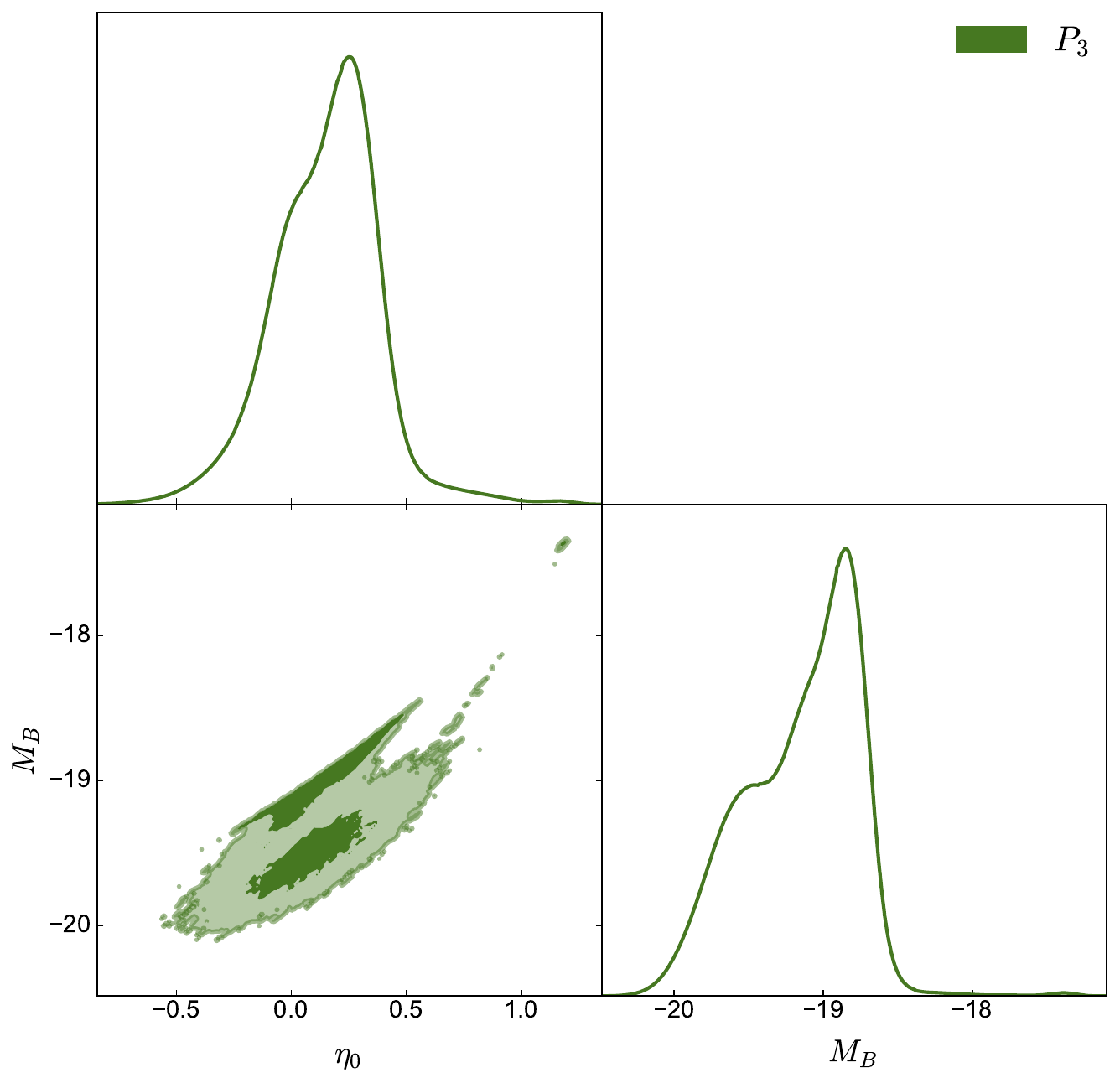}
\caption{Joint constraints on the DDR violation parameter $\eta_0$ and the absolute magnitude of Type Ia supernovae $M_B$ for three different parameterizations of $\eta(z)$. The panels show the two-dimensional contours (68\% and 95\% confidence levels) and one-dimensional marginalized posterior distributions for $P_1: \eta(z) = 1 + \eta_0 z$ (blue), $P_2: \eta(z) = 1 + \eta_0 \frac{z}{1+z}$ (purple), and $P_3: \eta(z) = 1 + \eta_0 \log(1+z)$ (green).}

\label{fig:fig2}
\end{figure*}

When $M_B$ is allowed to vary as a free parameter, we find interesting correlations between $\eta_0$ and $M_B$ for all three parameterizations. Figure \ref{fig:fig1} shows the two-dimensional contour plots and one-dimensional posterior distributions for $\eta_0$ and $M_B$ for each parameterization.

For the $P_1$ model, where $\eta(z) = 1 + \eta_0 z$, our analysis yields the following constraints: $\eta_0 = 0.20^{+0.11}_{-0.21}$ and $M_B = -18.86^{+0.21}_{-0.56}$ mag. The constraint on $\eta_0$ is consistent with zero at approximately 1$\sigma$ level, suggesting no strong evidence for a violation of the DDR. The constraint on $M_B$ is notably different from the commonly accepted value of approximately $-19.3$ mag. This shift could be attributed to the correlation between $M_B$ and $\eta_0$, as allowing for DDR violations affects the luminosity distance calculations and consequently the inferred absolute magnitude of SNe Ia.

For the $P_2$ model, where $\eta(z) = 1 + \eta_0 \frac{z}{1+z}$, we obtain the following constraints: $\eta_0 = 0.19^{+0.17}_{-0.23}$ and $M_B = -19.00^{+0.22}_{-0.56}$ mag. Similar to the $P_1$ model, the constraint on $\eta_0$ suggests no strong evidence for a violation of the DDR. The constraint on $M_B$ is closer to the commonly accepted value of approximately $-19.3$ mag compared to the $P_1$ model, but still shows a slight shift toward dimmer absolute magnitudes.

The similarity in the central values of $\eta_0$ between the $P_1$ and $P_2$ models ($0.2$ and $0.19$, respectively) suggests a degree of robustness in the detected trend toward a small positive DDR violation. However, the large error bars for both $\eta_0$ and $M_B$ across the two models indicate that this trend is not statistically significant. These results emphasize the importance of considering multiple parameterizations when testing for DDR violations, as different models can provide complementary insights into the nature of any potential deviations from the standard DDR.

For the $P_3$ model, where $\eta(z) = 1 + \eta_0 \log(1+z)$, our analysis yields the following constraints: $\eta_0 = 0.17^{+0.18}_{-0.23}$ and $M_B = -19.08^{+0.31}_{-0.49}$ mag. These results provide further insights into potential DDR violations and complement our findings from the $P_1$ and $P_2$ models. The constraint on $\eta_0$ remains consistent with zero within approximately 1$\sigma$, continuing the trend observed in the previous models.

Notably, all three parameterizations are consistent with $\eta_0 = 0$ within 1$\sigma$ uncertainties, providing support for the standard cosmological model and the validity of the DDR. However, the relatively large uncertainties in $\eta_0$ leave room for potential small deviations from the standard DDR. These results underscore the importance of considering multiple parameterizations in DDR violation tests. All three models point toward similar conclusions, demonstrating the robustness of our DDR violation detection capabilities.

The observed correlations between $\eta_0$ and $M_B$ have important implications for cosmological studies. They suggest that any potential violation of the DDR could affect our estimates of the intrinsic luminosity of SNe Ia, which in turn could impact measurements of cosmological parameters, including the Hubble constant. This interaction between DDR violations and standard candle calibrations underscores the need for careful consideration of these effects in precision cosmology.

An interesting feature of our results when $M_B$ is treated as a free parameter is the notably nonsmooth nature of the contours in the $\eta_0$--$M_B$ plane, as shown in Figure \ref{fig:fig2}. This irregularity is particularly striking and warrants further discussion. To ensure that the nonsmooth contours in the $\eta_0$--$M_B$ plane are not due to insufficient MCMC sampling, we performed extensive convergence tests. We tested the MCMC sampling with 10,000, 20,000, and 40,000 points. The persistence of the nonsmooth features across these tests, with consistent results regardless of the number of sampling points, confirms that they are inherent to the likelihood surface and not artifacts of the sampling process. This robustness check strengthens our confidence in the complex relationship observed between $\eta_0$ and $M_B$, suggesting it is a genuine feature of our model rather than a sampling artifact.

The multimodal structure in the $\eta_0$--$M_B$ plane may be fundamentally connected to the absolute distance calibrations in our analysis. The time-delay distance $D_{\Delta t}$ is a combination of three distances (the angular diameter distances between observer-lens, lens-source, and observer-source). The model prediction $\mu_{D_{\Delta t}}$ in our likelihood analysis is calibrated using SNe Ia distances, which critically depends on the absolute magnitude $M_B$. Without proper $M_B$ calibration, we cannot obtain absolute distances, and consequently cannot derive the model prediction for time-delay distance.

Meanwhile, the observational time-delay distance $D_{\Delta t}$ from strong lensing measurements also contains absolute distance information. When constructing the likelihood using the residuals between these two absolute distances, the multimodal structure could emerge if there exists some tension between these distance measurements. Specifically, the high-density regions in the posterior distribution might represent different possible reconciliations between the distance scales from SNe Ia and strong lensing observations.

The multimodal structure could indicate either a potential systematic difference in the distance scales derived from these two distinct observational probes, or a genuine deviation from the standard distance duality relation, as parameterized by $\eta_0$. We cannot definitively distinguish between these possibilities with current data alone.

The complexity in the $\eta_0$--$M_B$ posterior is therefore not merely a statistical feature but could be revealing important physical information about the consistency between different distance measurements in cosmology. This interpretation suggests that the observed structure warrants further investigation with additional independent distance indicators to better understand its physical origin.

\subsection{Constraints with Fixed SNe Ia Absolute Magnitude}
\begin{figure*}
\centering
\includegraphics[scale=0.29]{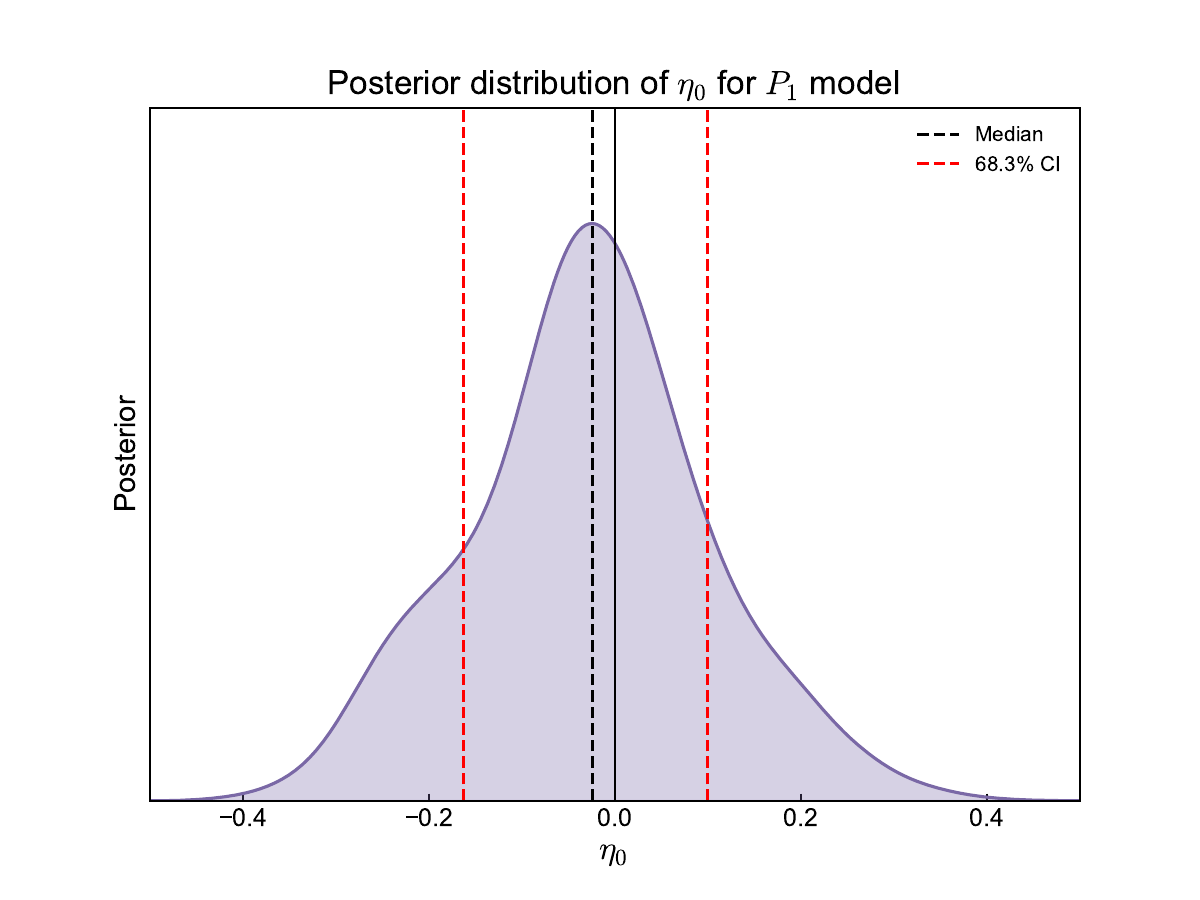}
\includegraphics[scale=0.29]{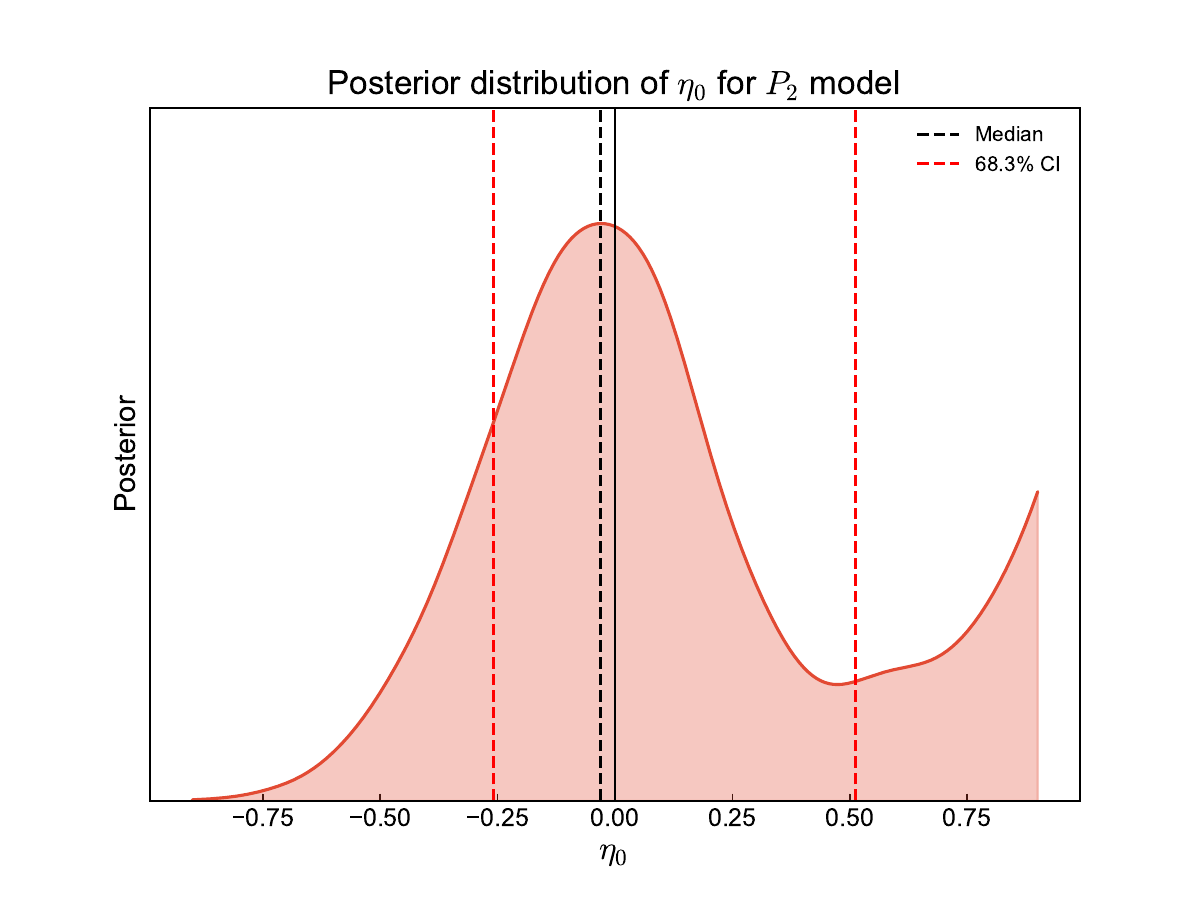}
\includegraphics[scale=0.29]{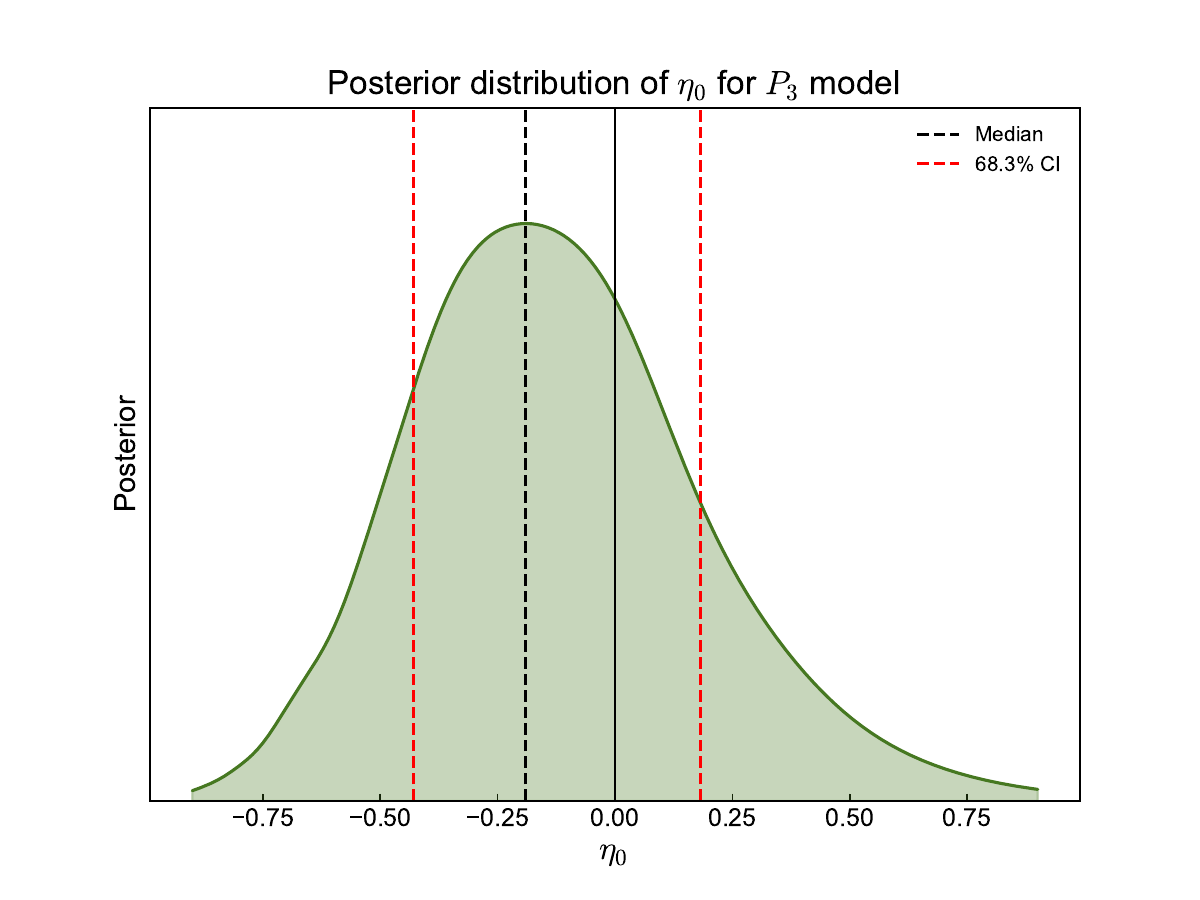}
\caption{Posterior distributions of the $\eta_0$ parameter for three different models of the DDR with $M_B$ fixed to $-19.253 \pm 0.027$ mag. The solid black line represents the median value, while the dashed red lines indicate the 68.3\% confidence interval.}

\label{fig:fig3}
\end{figure*}

To further investigate the robustness of our constraints on DDR violations, we performed a second analysis with $M_B$ fixed to $-19.253 \pm 0.027$ mag from the Cepheid-calibrated distance ladder method \citep{Riess:2021jrx}. Figure \ref{fig:fig3} shows the posterior distributions of $\eta_0$ for each of the three parameterizations under this fixed $M_B$ scenario. The results for each parameterization are $\eta_0 = -0.02 \pm 0.13$ for $P_1$, $\eta_0 = -0.03 ^{+0.23}_{-0.54}$ for $P_2$, and $\eta_0 = -0.19 ^{+0.24}_{-0.37}$ for $P_3$.

While $P_1$ and $P_2$ yield results very close to zero, $P_3$ shows a more pronounced negative value but is still consistent with zero within 1 $\sigma$ uncertainties. This suggests that the logarithmic form of $P_3$ might be more sensitive to potential DDR violations when $M_B$ is fixed. Notably, all three parameterizations yield results consistent with $\eta_0 = 0$ within 1$\sigma$ uncertainties, providing strong support for the validity of the standard DDR. This consistency across different models strengthens our confidence in the robustness of the DDR.

Interestingly, the central values of $\eta_0$ have changed from positive (when $M_B$ was free) to negative for all parameterizations. This sign flip highlights the strong degeneracy between $\eta_0$ and $M_B$, demonstrating how fixing $M_B$ can significantly impact our inferences about potential DDR violations. While fixing $M_B$ provides a more direct probe of DDR violations, it may also introduce biases if the fixed value is not precisely correct. The complementary information from both approaches provides a more comprehensive understanding of potential DDR violations and their cosmological implications.

A notable feature in our analysis is the increase in $\eta_0$ uncertainties when a prior is applied to $M_B$ (Table \ref{tab:results}). This behavior is closely tied to the structure of the $\eta_0$--$M_B$ parameter space, as revealed by our contour plots. These plots show two distinct high-density regions. When $M_B$ is allowed to vary freely, the best-fit value of $\eta_0$ falls within the upper, more compact high-density region, resulting in smaller uncertainties. This region corresponds to $M_B$ values of approximately $-19.0$ mag or slightly larger. However, fixing $M_B$ to $-19.253$ mag shifts the best-fit $\eta_0$ to the lower, more extended high-density region, leading to larger uncertainties. This shift also explains the change in $\eta_0$'s central value from positive to negative or near-zero when $M_B$ is fixed. These results highlight the complex interplay between $\eta_0$ and $M_B$, underscoring the importance of carefully considering priors in analyses of the distance duality relation.

\section{Conclusion}\label{sec:conclusion}

In this work, we present a comprehensive investigation of the cosmic DDR using a combination of SGL time delay measurements and SNe Ia data. By employing three different parameterizations of potential DDR violations, we test the validity of this fundamental relation in cosmology. Our analysis, considering both free and fixed absolute magnitude ($M_B$) scenarios for SNe Ia, yields several important conclusions:
\begin{itemize}
\item Across all three parameterizations ($P_1$, $P_2$, and $P_3$), we find no statistically significant evidence for violations of the DDR. The consistency of $\eta_0$ with zero within 1$\sigma$ uncertainty provides strong support for the standard DDR.

\item The observed correlations between the DDR violation parameter $\eta_0$ and the absolute magnitude $M_B$ of SNe Ia highlight the intricate relationship between potential DDR violations and the calibration of standard candles. The nonsmooth contours in the $\eta_0$--$M_B$ plane suggest a complex parameter space, potentially indicating subtle systematic effects or model dependencies.

\item The comparison between free and fixed $M_B$ scenarios reveals the sensitivity of DDR violation constraints to supernova calibration. The sign flip in $\eta_0$ when fixing $M_B$ emphasizes the need for careful consideration of absolute magnitude uncertainties in DDR tests.

\item The consistency of our results across different parameterizations demonstrates the robustness of our analysis methodology. 
\end{itemize}

While our findings strongly support the validity of the DDR, the uncertainties in our constraints leave room for small deviations. Future work could benefit from larger datasets, particularly at high redshifts, and improve modeling of systematic effects. Additionally, exploring alternative parameterizations and incorporating other cosmological probes could further refine our understanding of the DDR and its implications for fundamental physics.

In conclusion, our analysis provides strong support for the validity of the cosmic distance duality relation, with no significant evidence for violations across multiple parameterizations and analysis approaches. These results not only reinforce our confidence in the standard cosmological model but also demonstrate the power of combining different cosmological probes in testing fundamental physical principles. As we continue to gather more precise cosmological data, tests of the DDR will remain a crucial tool for probing the foundations of our understanding of the Universe.

\section*{Acknowledgments}
This work was supported by the National SKA Program of China (grants Nos. 2022SKA0110200 and 2022SKA0110203), the National Natural Science Foundation of China (grants Nos. 12473001, 12205039 and 11975072), the National 111 Project (grant No. B16009), and the China Manned Space Project (grant No. CMS-CSST-2021-B01). J.-Z. Q. is also funded by the China Scholarship Council.

\appendix

\section{Distance Conversion and Error Propagation}\label{app:distance_conversion}

In our analysis, we convert luminosity distances ($D_L$) obtained from supernova data to angular diameter distances ($D_A$) and time-delay distances ($D_{\Delta t}$) used in strong gravitational lensing. This appendix details the conversion process and error propagation.

\subsection{Luminosity Distance to Angular Diameter Distance}

We convert luminosity distance to angular diameter distance using the DDR with a possible violation parameter $\eta(z)$:
\begin{equation}
D_A(z) = \frac{D_L(z)}{(1+z)^2\eta(z)}.
\end{equation}
The error in $D_A$ is propagated as
\begin{equation}
\sigma_{D_A}(z) = \frac{\sigma_{D_L}(z)}{(1+z)^2\eta(z)},
\end{equation}
where $\sigma_{D_L}(z)$ is the error in the luminosity distance.

\subsection{Time-delay Distance Calculation}

For a strong lensing system with lens redshift $z_d$ and source redshift $z_s$, we calculate the time-delay distance $D_{\Delta t}$ as follows:

First, we compute the angular diameter distances to the lens ($D_d$) and source ($D_s$):
\begin{equation}
D_d = D_A(z_d), \quad D_s = D_A(z_s).
\end{equation}
Then, we calculate the ratio of the lens-source distance to the source distance:
\begin{equation}
\frac{D_{ds}}{D_s} = 1 - \frac{(1+z_d)D_d}{(1+z_s)D_s}.
\end{equation}
Finally, we compute the time-delay distance:
\begin{equation}
D_{\Delta t} = (1+z_d)\frac{D_d D_s}{D_{ds}}.
\end{equation}

\subsection{Error Propagation for Time-delay Distance}

The error in $D_{\Delta t}$ is calculated using error propagation:
\begin{equation}
\sigma_{D_{\Delta t}}^2 = \left(\frac{\partial D_{\Delta t}}{\partial D_d}\right)^2 \sigma_{D_d}^2 + \left(\frac{\partial D_{\Delta t}}{\partial D_s}\right)^2 \sigma_{D_s}^2,
\end{equation}
where
\begin{equation}
\frac{\partial D_{\Delta t}}{\partial D_d} = \frac{D_{\Delta t}}{D_d} + \frac{D_{\Delta t}^2}{D_d D_s (1+z_s)},
\end{equation}
\begin{equation}
\frac{\partial D_{\Delta t}}{\partial D_s} = \frac{D_{\Delta t}}{D_s} - \frac{D_{\Delta t}^2}{D_d D_s (1+z_d)}.
\end{equation}
These equations allow us to propagate the uncertainties from the luminosity distances to the time-delay distances used in our analysis of strong gravitational lensing systems.

\section{Incorporating Supernova Data Errors in Likelihood Calculations} \label{app}

To account for the uncertainties introduced by the supernova (SN) data when testing the cosmic DDR, we have modified several likelihood functions. This appendix provides a detailed description of these modifications.

\subsection{Skewed Log-normal Likelihood}

For the skewed log-normal likelihood function, we modified the original function to include the error from the SN data ($\sigma_{D_L}$) for the time-delay distance ($D_{\Delta t}$):

\begin{equation}
\mathcal{L}_{D_{\Delta t}} = \frac{1}{\sqrt{2\pi} (D_{\Delta t} - \lambda) \sqrt{\sigma^2 + \sigma_{D_L}^2}} \exp \left[ -\frac{(\mu - \log(D_{\Delta t} - \lambda))^2}{2(\sigma^2 + \sigma_{D_L}^2)} \right].
\end{equation}

where $\mu$, $\sigma$, and $\lambda$ are the parameters of the skewed log-normal distribution.

\subsection{Joint Skewed Log-normal Likelihood}

For the joint skewed log-normal likelihood function for the angular diameter distance ($D_A$) and the time-delay distance ($D_{\Delta t}$), we included the errors from the SN data for both quantities:

\begin{equation}
\mathcal{L}_{D_A} = \frac{1}{\sqrt{2\pi} (D_A - \lambda_{D_A}) \sqrt{\sigma_{D_A}^2 + \sigma_{D_A, \text{SN}}^2}} \exp \left[ -\frac{(\mu_{D_A} - \log(D_A - \lambda_{D_A}))^2}{2(\sigma_{D_A}^2 + \sigma_{D_A, \text{SN}}^2)} \right],
\end{equation}

\begin{equation}
\mathcal{L}_{D_{\Delta t}} = \frac{1}{\sqrt{2\pi} (D_{\Delta t} - \lambda_{D_{\Delta t}}) \sqrt{\sigma_{D_{\Delta t}}^2 + \sigma_{D_{\Delta t}, \text{SN}}^2}} \exp \left[ -\frac{(\mu_{D_{\Delta t}} - \log(D_{\Delta t} - \lambda_{D_{\Delta t}}))^2}{2(\sigma_{D_{\Delta t}}^2 + \sigma_{D_{\Delta t}, \text{SN}}^2)} \right].
\end{equation}

\subsection{Gaussian Likelihood}

For the Gaussian likelihood function, we modified it to include the error from the SN data:

\begin{equation}
\mathcal{L}_{D_{\Delta t}} = \frac{1}{\sqrt{2\pi (\sigma_{D_{\Delta t}}^2 + \sigma_{D_{\Delta t}, \text{SN}}^2)}} \exp \left[ -\frac{(D_{\Delta t} - \mu_{D_{\Delta t}})^2}{2(\sigma_{D_{\Delta t}}^2 + \sigma_{D_{\Delta t}, \text{SN}}^2)} \right].
\end{equation}

\subsection{Kernel Density Estimator (KDE) Likelihood}

For likelihood functions using KDE, we adjusted the bandwidth to account for the SN data errors:

\begin{equation}
\text{effective bandwidth} = \sqrt{\text{original bandwidth}^2 + \sigma_{\text{SN}}^2}.
\end{equation}

This modification was applied to both the full KDE likelihood and the histogram-based KDE likelihood.

\subsection{Histogram Linear Interpolation Likelihood}

For the histogram linear interpolation likelihood, we incorporated the SN data error by adding it to the total variance:

\begin{equation}
\log \mathcal{L}_{D_{\Delta t}} = -\frac{1}{2} \left[ \frac{(D_{\Delta t} - \text{interpolated value})^2}{\sigma_{D_{\Delta t}, \text{SN}}^2} + \log(2\pi \sigma_{D_{\Delta t}, \text{SN}}^2) \right].
\end{equation}

\subsection{General Likelihood}

For a general likelihood function, we included the SN data error by adding it to the total uncertainty:

\begin{equation}
\mathcal{L}_{D_{\Delta t}} = -\frac{(D_{\Delta t} - \mu_{D_{\Delta t}})^2}{2(\sigma_{D_{\Delta t}}^2 + \sigma_{D_{\Delta t}, \text{SN}}^2)}.
\end{equation}

These modifications ensure that the errors from the SN data are properly accounted for in our likelihood calculations when testing the cosmic distance duality relation, providing a more accurate and reliable assessment of the DDR validity.

%\appendix

\bibliography{main}{}
\bibliographystyle{aasjournal}

\end{document}